\documentclass[aps,floats,epsf, twocolumn ]{revtex4}
\usepackage{amsfonts}
 \usepackage{graphics}
 
\begin{document}

\title{Inhomogeneous magnetic catalysis on graphene's honeycomb lattice}

\author{Bitan Roy and Igor F. Herbut}

\affiliation{Department of Physics, Simon Fraser University,
 Burnaby, British Columbia, Canada V5A 1S6}

\begin{abstract}
We investigate the ordering instability of interacting (and for simplicity, spinless) fermions on graphene's honeycomb lattice by numerically computing the Hartree self-consistent solution for the charge-density-wave order parameter in presence of both uniform and non-uniform magnetic fields. For a uniform field the overall behavior of the order parameter is found to be in accord with the continuum theory. In the inhomogeneous case, the spatial profile of the order parameter resembles qualitatively the form of the magnetic field itself, at least when the interaction is not overly strong. We find that right at the zero-field critical point of the infinite system the local order parameter scales as the square-root of the local strength of the magnetic field, apparently independently of the assumed field's profile. The finite size effects on various parameters of interest, such as the critical interaction and the universal amplitude ratio of the interaction-induced gap to the Landau level energy at criticality are also addressed.

\end{abstract}
\maketitle

\vspace{10pt}

\begin{center}
\section{Introduction}
\end{center}

Graphene has since its successful fabrication \cite{Geim} emerged as the prime electronic system of reduced dimensionality. Its structure can be described as two interpenetrating
triangular sublattices of carbon atoms, which together form a bipartite honeycomb lattice.
As a consequence of the lack of inversion symmetry around a site of honeycomb lattice, the valence
and the conduction bands touch each other at the six corners of the first Brillouin zone. At low
energies, one can linearize the electronic dispersion relation near those ``Dirac" points. In
particular, at the filling one half when the conduction band is empty and the valence band is
filled, gapless quasiparticle excitations live in the vicinity of the Dirac points. In the continuum
limit, such excitations can then be described in terms of pseudo-relativistic massless Dirac fermions,
with the Fermi velocity $v_F(\approx c/300)$ playing the role of the velocity of light ($c$) \cite{castro}. \\

In its usual state graphene behaves like a semi-metal. A large overlap of the electronic wave
functions of the neighboring carbon atoms ($t \sim 2.5$ eV)  protects such a phase against weak
electron-electron interactions. In the language of renormalization group, such a stability corresponds
to a large domain of attraction of the non-interacting Gaussian fixed point \cite{Igor}. On the other hand,
a strong enough interaction can bring on a Mott-Hubbard transition towards a gapped insulating phase
\cite{gonzales, bitan}. For example, a sufficiently strong on-site Hubbard interaction $(U)$ or nearest-neighbor
Coulomb repulsion $(V)$ would turn the system into an insulator with  the staggered pattern of either
average magnetization or density. For graphene \cite{Wehling}, $U \sim 10$ eV, $V\sim
2-5$ eV, whereas the critical values for insulation are $U_c/t \approx 4-5$ \cite{Sorella, Paiva} and $V_c/t
\approx 1 $ \cite{Franz}. It appears that graphene lies safely on the semi-metallic side of possible
Mott transitions, but with the interactions which are nevertheless not too far from their critical values.
Currently, the interaction-to-bandwidth ratios that control the Mott transitions
in graphene are not easily tunable. However, subjecting the system to a finite magnetic flux quenches the kinetic energy and collapses the density
of states (DOS) onto a discrete set of Landau levels (LLs), and can this way ``catalyze" the formation of ordered phases \cite{gusynin1}. In presence of a magnetic field even an infinitesimal amount of the on-site Hubbard $U$ or the nearest-neighbor repulsion $V$ would  turn the system at half filling into  an ordered phase with either finite
N$\acute{e}$el order or staggered density \cite{dima, herbut4}. Yet another and a qualitatively different insulator \cite{haldane} may result from  a strong next-nearest-neighbor repulsion, which can induce a gapped phase with finite circulating currents between the sites on the same sublattice \cite{raghu}. Such a phase, in the spinless case, violates the time-reversal symmetry and represents an early example of a topological insulator. The same topological insulator may also be possible to catalyze by a fictitious magnetic field \cite{herbut3} that would arise from specific deformations of the graphene sheet \cite{guinea, levy}, for example. \\

\begin{figure}[t]
{\centering\resizebox*{80mm}{!}{\includegraphics{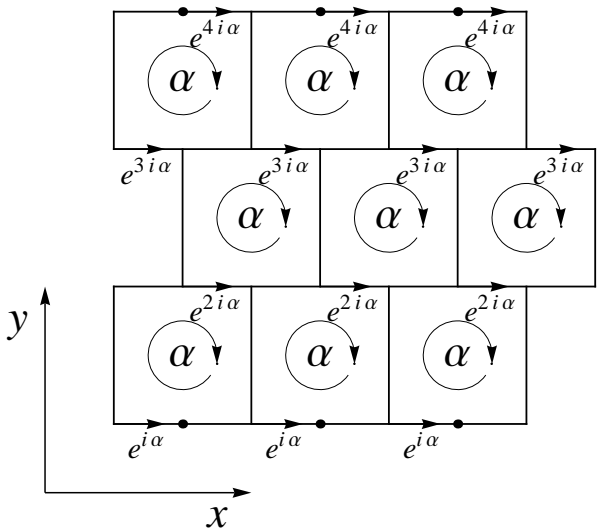}}
\par} \caption[] {Brickwall realization of honeycomb lattice. Here the magnetic flux of $\alpha \Phi_0$ pierces through each
hexagon, corresponding to a uniform magnetic field through the system. This particular choice of gauge
is equivalent to the Landau gauge  $\emph{A}=(-By,0)$ in the continuum description.
This construction is straightforward to generalize to an inhomogeneous magnetic field. }\label{brickwall}
\end{figure}

The magnetic catalysis in the presence of uniform magnetic field is by now well understood. It has been
proposed as a mechanism  behind the formation of the Hall states in graphene at filling factors $\nu=0$
and $\nu=1$ \cite{herbut4, gusynin}, which become discernible at  higher magnetic fields. \cite{Zhang}
Sublinear scaling of the gap with the magnetic field, for example, strongly suggests that the electron-electron
interactions are the cause of the gap at $\nu=1$ \cite{Zhang2, bitan2}. In contrast, the behavior of
interacting electrons in presence of an \emph{inhomogeneous} magnetic field has not been studied much, although the issue of order parameter's dependence on the local value of the magnetic field has been addressed analytically, for specific spatial profiles of the field \cite{Dunne, Raya}. In this work we attempt to develop  a more detailed  understanding of the spatial variation and the field dependence of the order parameter when an inhomogeneous magnetic field penetrates through the system. The motivation for such a study comes in part from a closely related problem of interacting electrons in a pseudomagnetic field \cite{herbut3, guinea}, where field's profile is typically non-uniform in space. On a methodological level, it seems also interesting to inquire how much of the catalysis mechanism remains in effect when the condition of uniformity of the magnetic field is relaxed.

A self-consistent profile of the local gap in the insulating phase is computed therefore numerically on a discrete lattice and at the level of Hartree approximation, with a special attention given to its spatial variation. We find that the system stills suffers a metal-insulator transition at weak nearest-neighbor interactions even in the presence of a inhomogeneous magnetic field. In the continuum, this phenomenon would be attributed to the delta-function density of states when the Fermi energy is at the Dirac point \cite{casher}. Interestingly, we find that the spatial profile of the interaction-induced gap (order parameter) in presence of a localized magnetic flux, although not matching exactly, still mimics closely the profile of the local strength of the magnetic field. Moreover, right at the zero-field metal-insulator quantum criticality, the {\it local}  order parameter seems to vary very much like the square-root of the {\it local} magnetic field. This behavior is analogous to what we previously found  analytically in a uniform magnetic field \cite{bitan2}, and to what we also confirm here numerically (see below). Away from the critical point, at weak interactions the expectation value of the local order parameter reverts to a linear dependence of the local magnetic field, as one might expect.
\\

 For a uniform magnetic field, we in general find a very good agreement between the previous field theoretic results and our numerical calculations. We focus on the finite-ranged components of the Coulomb repulsion, and chose to keep only the simplest one, which acts between the nearest neighbors. The system at the filling one-half and in the magnetic field then develops a gap in the spectrum, even when the interaction is weak.  Right at the metal-insulator quantum critical point the gap behaves as
\begin{equation}
m=\frac{E(1)}{C},
\end{equation}
where $E(n)$ is the $n^{th}$ LL energy. Here $C$ is the universal number,  found to be $6.1$ in the largest
system considered here, within the the Hartree approximation. This is in satisfactory agreement with
the same quantity computed previously in the field-theoretic description \cite{bitan2}, where we found it to
be $5.985$, in the limit of infinite number of fermion components. These two procedures being equivalent, we indeed find that upon increasing system's size the constant $C$ slowly approaches its value in the continuum.    \\

The rest of the discussion is organized as follows. The system of free electrons on honeycomb lattice in presence of magnetic field is introduced in Sec. II. The Hartree mean-field theory of the electrons interacting via the
nearest-neighbor repulsion is outlined in Sec. III. In that section we also demonstrate the mechanism of magnetic catalysis in a finite size system. Sec. IV focuses on the scaling behavior of the gap with the strength of the
uniform magnetic field and the interaction. Sec. V is devoted to the spatial variation of the gap in presence of nonuniform magnetic field. In Sec. VI, we summarize the results and discuss some related issues. \\

\begin{figure}[t]
{\centering\resizebox*{80mm}{!}{\includegraphics{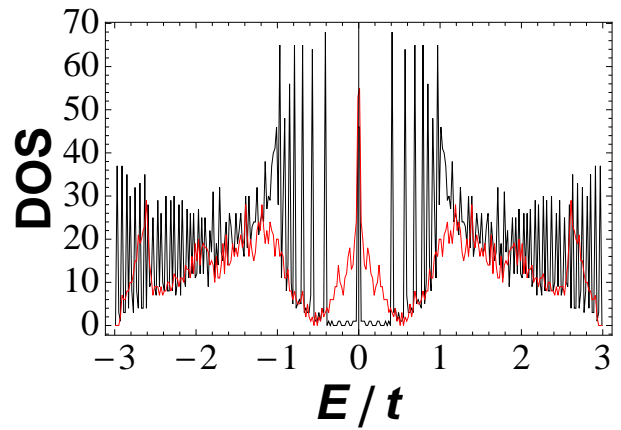}}
\par} \caption[] {DOS as a function of energy in presence of uniform (black) and non-uniform
(red) magnetic field. First five LLs are well formed in the case of a uniform magnetic field. In contrast, the DOS
is sharply peaked only at zero energy in presence of an inhomogeneous magnetic field.}\label{DOS}
\end{figure}

\section{Free fermions}

\begin{figure}[t]
{\centering\resizebox*{80mm}{!}{\includegraphics{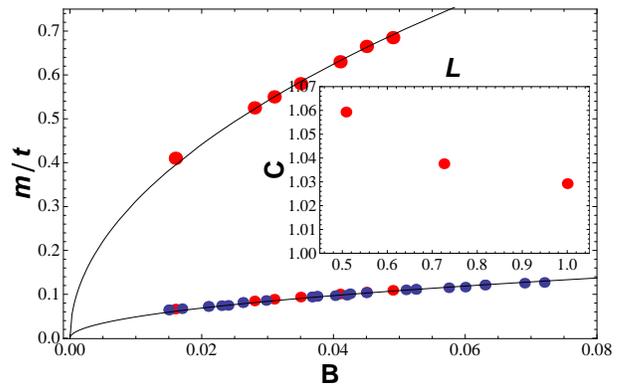}}
\par} \caption[] {Top curve corresponds to the first LL energies at various magnetic fields B (measured in the unit of $B_0$) in a finite lattice.
The energy spectrum is computed in a system of $40 \times 31$ points with a periodic boundary in x-direction.
The bottom one shows the interaction induced gap, as a function of magnetic field at the zero-field metal-insulator critical point $V/t=0.75$, in the same system. The red dots correspond to the OPs for the uniform field, whereas
the blue ones correspond to that in an inhomogeneous field, at different regions in the bulk of the system. Inset
shows the variation of the universal ratio ($C$) relative to its field-theoretic value ($5.985$) with the system size. Here $L$ corresponds to the ratio of system size to the maximum one. The largest lattice considered here has
$40\times 31$ lattice points.}\label{LLS}
\end{figure}

Let us define the system of non-interacting electrons on honeycomb lattice in the
presence of a uniform magnetic field. The tight-binding model with nearest-neighbor hopping is defined
as
\begin{equation}
H_t= -t \sum_{A,i} {c^{\dagger}_{A}}  {c_{A+b_{i}}} + H.c.,
\end{equation}
where $c$ and $c^\dagger$ are the usual fermionic annihilation and creation operators, respectively. Here we
omitted the spin degrees of freedom, for simplicity. $\vec{A}$ denotes the sublattice generated by the linear
combination of basis vectors $\vec{a}_1=(\sqrt{3},-1)a$ and $\vec{a}_2=(0,1)a$, for example. The second sublattice is then at $\vec{B}=\vec{A}+\vec{b}$, with $\vec{b}$ being either $\vec{b}_1=(1/{\sqrt{3}},1)a/2,\vec
{b}_2=(1/{\sqrt{3}},-1)a/2$ or $\vec{b}_3=(-1/{\sqrt{3}},0)a$, where $a$ is the lattice spacing. The magnetic
field may be introduced through the Peierls substitution $t \rightarrow {t} e^{(i 2\pi e / h)\int \vec{\emph{A}}
\cdot d\vec{l}}$, where $h/e =\Phi_0$ is the usual flux-quantum, and $(1/ \Phi_0) \int \vec{\emph{A}}\cdot d\vec
{l}=\Phi/\Phi_0$ counts the magnetic flux through each plaquette of the honeycomb lattice. In case of graphene, $\Phi_0$ corresponds to a magnetic field $\sim 10^4$ T, with commonly assumed lattice constant,
$a\approx 3$ $\mathring{A}$. Such a high magnetic field corresponds to the magnetic length close to the lattice scale, $B_0 \sim 1/a^2$. Therefore $\Phi/{\Phi_0}$ is equivalent to $B/B_0$, where $B/{B_0}=0.05$ corresponds to $B=500$ T.
In Fig.\ [\ref{brickwall}] we have shown one way to introduce a uniform magnetic field on  honeycomb lattice. By solving numerically the tight-binding model on a $80 \times 65$ lattice with periodic boundary in x-direction and the  field $B=160$ T, we clearly see the first few $(5)$ LLs as the well-separated energies  where the
DOS is sharply peaked (black curve Fig.\ [\ref{DOS}]). The energy spectrum is symmetric about
zero and the spacing among the LLs decreases with the LL index. The energy of the LLs varies as the square root of the
magnetic field, due to the relativistic nature of the quasi-particles (top curve Fig.\ [\ref{LLS}]). We also found that the maximum energy of the free electron system is $2.97 t (< 3t)$, in agreement with the previous results \cite{kohmoto}. It may be worth mentioning that in presence of the periodic boundary conditions the choice of gauge requires some care, and $\emph{A}(\vec{r})$ is chosen here so that only one out of the three bonds emanating  from a site contributes to it. Such a choice is then equivalent to the Landau gauge in the continuum description. \\

Next, we consider still non-interacting electrons on the honeycomb lattice, but now subject to an inhomogeneous  magnetic field. Numerically diagonalizing the free electron Hamiltonian, the DOS is found to be a smooth
function of energy, and peaked only at the zero energy (red curve in Fig.\ [\ref{DOS}]). There we considered a $70
\times 55$ lattice, with a open boundary and total flux $\Phi_{total}=4.59 \Phi_0$. We assumed the field to be uniform in the x-direction, and bell-shaped in y-direction, with the maximum at the center. The number of near zero energy states is proportional to the total flux of the magnetic field enclosed by the system.\cite{casher} The maximum energy in the free electron spectrum is found to be $2.93 t (< 3t)$. \\

\section{Interactions and magnetic catalysis}

Next, we turn on the short-range electron-electron interaction. The Hamiltonian in the
presence of only the nearest-neighbor Coulomb repulsion $(V)$ is given by
\begin{equation}
H=H_{t}+ \frac{V}{2} \sum_{\langle i,j \rangle }n_i n_j -\mu N,
\end{equation}
where $\langle i,j\rangle $ stands for the summation over the nearest-neighbor sites, $N$ is the total
number of electrons and $\mu$ is the chemical potential. After the usual Hartree decomposition the
effective single-particle Hamiltonian for interacting electrons becomes
\begin{eqnarray}
H _{HF} &=& H_{t}+ {V} \sum_{{<i,j>}} ( <n_{B,j}> n_{A,i}  \nonumber \\
&+& <n_{A,j}> n_{B,i} )
  -\mu N,
\end{eqnarray}
where $\langle n_{B(A)}\rangle $ counts the self-consistent site-dependent
average electron density on sub-lattice B(A). Let us measure these relative
to  the uniform density at half-filling by defining
\begin{equation}
<n_{A,i}> =\frac{1}{2}+\delta_{A,i}, \quad <n_{B,i}>=\frac{1}{2}-\delta_{B,i}.
\end{equation}
The positive quantities $\delta_A,\delta_B$ determine the local charge-density-wave order parameter (OP).
Both $\delta_A$ and $\delta_B$ will be functions of position with the constraint that the system is precisely
at half filling,
\begin{equation}
\sum_{i}\delta_{A,i} -\sum_{i}\delta_{B,i}=0.
\end{equation}
We also choose the value of $\mu= V/2$.
\\

\begin{figure}[t]
{\centering\resizebox*{80mm}{!}{\includegraphics{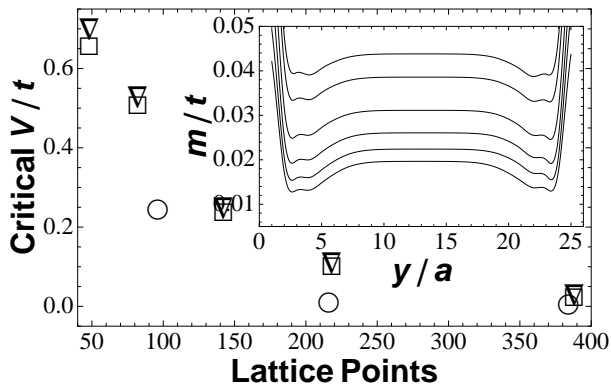}}
\par} \caption[] {The variation of finite-size ``critical interaction" $(V/t)_C$ (see the text)
as a function of the lattice size.
$\bigtriangledown$ and $\Box$ stand for $(V/t)_C$ in a cylindrical lattice with open boundary in
different gauges $\emph{A}=(0,Bx)$ and $\emph{A}=(-By,0)$, respectively. $\bigcirc$ stands for
$(V/t)_C$ in a lattice with open boundary, preserving the $C_6$ symmetry of the honeycomb lattice,
with $\emph{A}=(0,Bx)$. Here the critical interactions are computed for $\Phi/\Phi_0=0.05$. Inset
shows the variation of the size of the order parameter $(\delta_R)$ as a function of $V/t$, in the
entire system, computed on a $36 \times 25$ lattice. $V/t$ reads as $0.5,0.4,0.3,0.2,0.1,0.05$ from
top to bottom. } \label{criticalinteraction}
\end{figure}

We have computed the (Hartree) self-consistent solutions for the OPs, for different values of the flux and for
a variety of interaction strengths ($V/t$), at $T=0$. Consider a lattice with a periodic boundary in the $x$
direction and let us conveniently define the local OP as,
\begin{equation}
\delta_R=\frac{1}{2}(\delta_A+\delta_B), \label{OP}
\end{equation}
where $B$ is either one of the two nearest-neighbors to the site $A$, on the same row in x-direction. $\delta_R$ this way measures the order parameter
in a unit cell. On the other hand, the OP will be averaged over the points connected by the $C_6$ symmetry,
when we considered a quasi-circular system with an open boundary. In the presence of a uniform magnetic field,
$\delta_R$ is found to be uniform in the bulk of the system. However, $\delta_R$ becomes position dependent and
proportional to the local field when the system is subject to a inhomogeneous field.
\\

\begin{figure}[t]
{\centering\resizebox*{70mm}{!}{\includegraphics{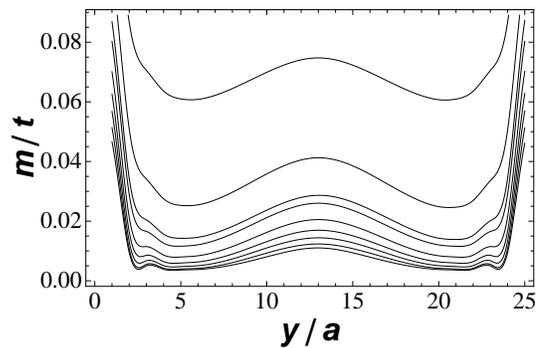}}
\par} \caption[] {Spatial distribution of the order parameter in the presence of a localized flux of magnetic field, with
total flux $\Phi_{total}=5.6 \Phi_0$.
The top curve corresponds to $V/t=0.76$ and the rest read as $V/t=0.65, 0.55, 0.5, 0.4, 0.3, 0.2, 0.1, 0.05, 0.01$
from top to bottom. }\label{critinhomo}
\end{figure}

Before we proceed, it is worth pausing to establish a practical definition of the ``critical interaction" associated with the metal-insulator transition in the finite size system like ours. For a sufficiently strong nearest-neighbor interaction, fermions reside only on one sublattice, with the other one completely empty. The system is then deep in the insulating phase. As the interaction is weakened, the size of the order parameter decreases and we numerically find the system to go through a well-defined transition into the semi-metallic phase, where the
order parameters $\delta_A, \delta_B$ are zero in the entire system. Right above that particular interaction $(V/t)$ which we call critical, there is a finite, but slightly inhomogeneous staggered density {\it everywhere} in the system. We will designate that value as the critical interaction ${(V/t)}_C$ corresponding to the metal-insulator transition. The described scenario is quite generic and occurs both in the absence and presence of magnetic fields, which also may be either uniform or nonuniform. The observed non-analytic behavior in a finite system, however, is clearly an artifact of the Hartree approximation, i. e. of the requirement of the self-consistency of the solution. \\

We computed the variation of the critical interaction defined this way
with the lattice size and the geometry, for a particular
magnetic flux $\Phi/{\Phi_0}=0.05$  ($B=500 T$) through each plaquette of the lattice. As may be seen from the Fig.\ [\ref{criticalinteraction}], for a small system size the critical interaction is large even in the presence
of a magnetic field, and also depends on the geometry of the lattice, as well as on the choice of the gauge. Upon
increasing the size of the system, the value of the critical interaction decreases and appears to approach \emph{zero}
in the thermodynamic limit, in agreement with the results obtained in the continuum theory. A typical distribution of
the OP in a lattice with periodic boundary is shown in inset of Fig.\ [\ref{criticalinteraction}]. From that one can
conclude that upon decreasing $(V/t)$, the size of the order parameter decreases in the entire system, both
in the bulk and at the edge. However a finite gap in the spectrum exists even at a rather weak  interaction, $V/t=0.05$ (bottom curve). Hence, in the presence of a uniform magnetic field, we expect that a large system would find itself in a gapped insulating phase even at an infinitesimal interaction, as found in the continuum theory.
\\

On the other hand, when the system is exposed to a localized flux of magnetic field
OP develops a local expectation value (see Fig.\ [\ref{critinhomo}]). The local OP is found to be proportional to the local magnetic field. As one enters the regime of weaker interaction OP decreases both in the bulk and the edge of the system. Yet we managed to observe finite expectation value of the OP in the entire system even at the smallest interaction considered here, $V/t=0.01$. Therefore, the system can also find itself in a ordered phase at weak  interaction when an \textit{inhomogeneous} flux of the magnetic field pierces through it. This phenomenon can be attributed to the finite density of state at zero energy, where the chemical potential lies at the filling one-half. \\

\begin{figure}[t]
{\centering\resizebox*{80mm}{!}{\includegraphics{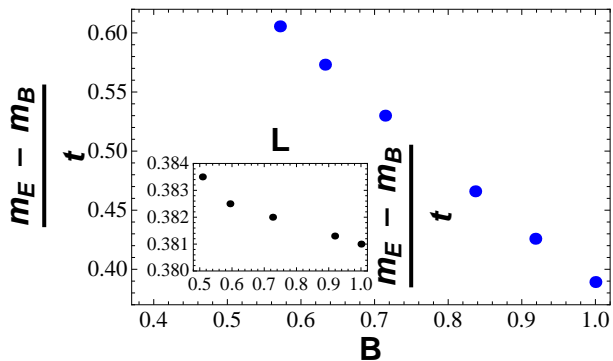}}
\par} \caption[] {Difference of the masses at the edge ($m_E$) and in the bulk ($m_B$) as a function of
the magnetic field ($B/B_{max}$) in a $36 \times 25$ lattice, at $V/t=0.5$. Inset: same quantity as a function
of system size ($L$), at fixed magnetic field $B/B_0=0.028$, at $V/t=0.76$. Here $B_{max}=490 T$ and $L_{max}=40
\times 31$. }\label{finite}
\end{figure}

Besides the condensation in the bulk of the system, the OP acquires spikes near
the edges of the system, in presence of both uniform (Inset of Fig.\ [\ref{criticalinteraction}]) and non-uniform (Fig.\ [\ref{critinhomo}]) field. The spikes in the OP near the edge of the system arise from the finite size effect. Such edge effects die out as one increases
the system size. Moreover, with the increasing magnetic field, those spikes also dissolve and give rise to a uniform condensation throughout the system, at sufficiently large magnetic field. These effects on the OP are demonstrated in Fig.\ [\ref{finite}]. We exhibited the finite-size effects in presence of a uniform flux only, but the result is qualitatively the same in presence of a localized flux as well.
\\

\section{Scaling in uniform magnetic field}

\begin{figure}[t]
{\centering\resizebox*{80mm}{!}{\includegraphics{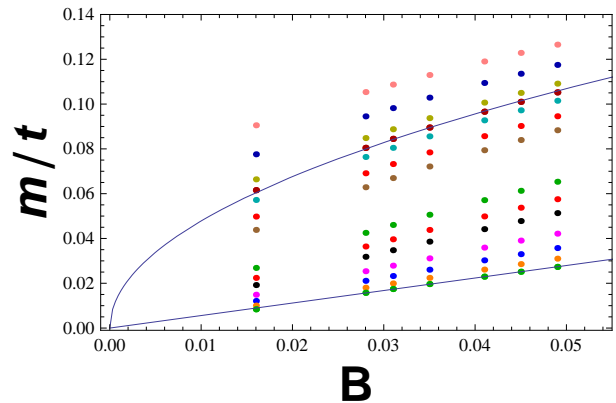}}
\par} \caption[] {OP ($\delta_R $ or m) as a function of $\Phi/\Phi_0$ or $B/B_0$ at different $V/t$. The top points
corresponds to $V/t=0.8$ and the remaining ones to $V/t=0.78,0.76,0.75,0.74,0.72,0.7,0.6,0.55,0.5,0.4,0.3,0.2,0.1$
from top to bottom. For $V/t=0.75$ we found the best $\sqrt{B}$ fit of the mass with magnetic field. }\label{MB}
\end{figure}

We now investigate the dependence of the gap on the magnetic field $(B)$ and the interaction $(V/t)$.
First we take the field to be uniform, and still consider the lattice with the periodic boundary in the $x$ direction.
The functional dependence of $\delta_R$, defined in Eq.\ \ref{OP}, on the magnetic fields and interactions is shown in Fig.\ [\ref{MB}]. Here, $\delta_R$ is computed on a $36 \times 25$ lattice and we considered OP only in the bulk of the system. With the parametrization as in Sec. II, the lowest value of the magnetic field ($160$ T), considered here, is about four times larger than the current highest constant laboratory magnetic field. However, upon using a larger system one can get down to a more realistic strength of the field. \\

For sufficiently small interactions, order parameter ($\delta_R$ or $m$) varies almost linearly with $B/B_0$ (bottom
curve in Fig.\ [\ref{MB}]). As one increases the strength of interaction, there is a  crossover to a sub-linear dependence
of the mass ($m$) on $B$ \cite{bitan2}. In particular, right at $V/t=0.75$, we find the best overall $\sqrt{B}$ fit of the mass to the magnetic field. We therefore designate that interaction to be the critical interaction $(V/t)_C$ at $B=0$. When we computed the ratio of the first LL energy to the interaction induced gap $(m)$  at $V/t=0.75$, it came out to be a universal number $(C) \approx
6.21$, independent of the magnetic field $B$ (inset Fig.\ [\ref{LLS}]). The value of the  number is in satisfactory agreement with the same quantity previously calculated in the continuum description and in the large-N limit \cite{bitan2}. There we obtained $C=5.985$, with the difference between the two values that can be attributed to the finite size (Inset Fig.\
[\ref{LLS}]). \\

We also found a similar dependence of the mass on interactions and magnetic fields, computed on a quasi-circular lattice with open boundary, preserving the $C_6$ symmetry of a hexagon. A spatial variation of the interaction induced gap in the presence of a uniform magnetic field is shown in Fig.\ [\ref{rotsymm}] (black curves). In that case, we found the
ratio of the first LL energy to the interaction induced gap $(C)$ to be $6.29$, in a system of $384$ lattice points. Such a particular choice of lattice turns out to be useful when one imposes rotationally symmetric inhomogeneous magnetic field. By considering a graphene sheet with open boundary, we computed the OP in two different gauges, equivalent to
$\emph{A}=(0,Bx)$ and $\emph{A}=(-By,0)$ and it turned out to be gauge independent, as expected.  \\

\section{Interacting fermions in inhomogeneous field}

Next, we consider spinless interacting fermions on a honeycomb lattice subject to a inhomogeneous magnetic field in more detail. It was previously shown that, for a specific realization of the inhomogeneous magnetic field and in the limit of a large magnetic flux, the order parameter in the insulating phase computed within the zero-energy manifold matches exactly the local profile of the magnetic field  \cite{Dunne}. Here we determine the order parameter self-consistently (at $T=0$) and on the honeycomb lattice, and include the contributions from all the states into account. We will consider two specific configurations of spatially modulated magnetic field. $(a)\quad$ Localized field in one direction,
$y$ in our case, but extended in the orthogonal direction, and $(b) \quad$ rotationally symmetric localized field with the maximum strength at the center. We imposed the field of type $(a)$ on a lattice with periodic boundary in $x$ direction. On
the other hand, a quasi-circular lattice with open boundary, preserving the $C_6$ symmetry of a hexagon, is exposed to a localized field of type $(b)$. As mentioned previously, even in presence of a non-uniform field, there is a large (and in the continuum limit, infinite) DOS at  zero energy. Therefore, right at the filling one half, one expects that even a weak interaction can place the system into a insulating phase. \\

\begin{figure}[t]
{\centering\resizebox*{80mm}{!}{\includegraphics{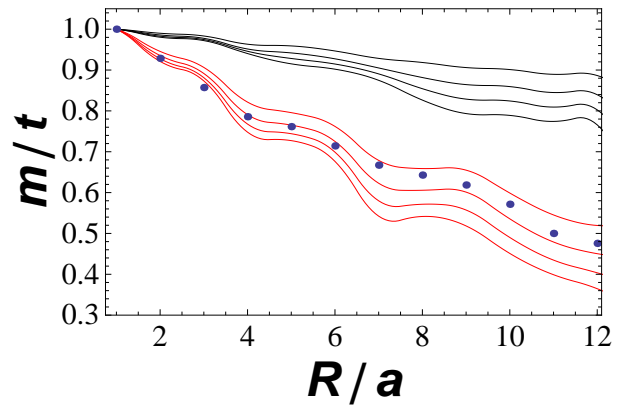}}
\par} \caption[] {Normalized OP in presence of a uniform (black) and a non-uniform (red) magnetic field for different values of $V/t$. From top to bottom $V/t$ reads  $0.6,0.5,0.4,0.3$. For the inhomogeneous field the total flux through the system is $\Phi_{total}=7.1 \Phi_0$. The blue dots corresponds to the local strength of the inhomogeneous magnetic field. }
\label{rotsymm}
\end{figure}

\begin{figure}[t]
{\centering\resizebox*{80mm}{!}{\includegraphics{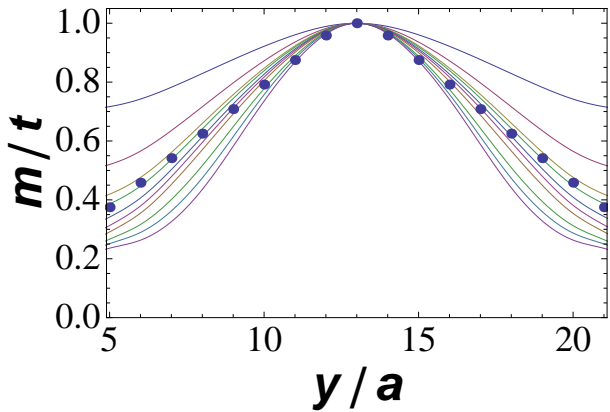}}
\par} \caption[] {Normalized OP in presence of inhomogeneous magnetic field at $\Phi_{total}=9.86 \Phi_0$. Average
magnetic field at different rows is denoted by the dots. $V/t$ reads  $0.75,0.65,0.55,0.5,0.4,0.3,0.2,0.1,0.05,
0.01$ from top to bottom. The blue dots corresponds to the local strength of the inhomogeneous magnetic field.}
\label{IHMC}
\end{figure}

In the presence of a non-uniform field, but with a finite total magnetic flux, the system develops a gap in the spectrum even at sub-critical interactions ($V/t \ll (V/t)_c$) (Fig.\ [\ref{critinhomo}]). The spatially variation of the order parameter $(\delta_R)$ in presence
of a inhomogeneous magnetic field of type $(a)$ is depicted in Fig.\ [\ref{IHMC}]. We considered the OP
only far from the edges of the system, and normalized the OP as well the magnetic fields with respect to their
maximum values at the center of the system. The order parameter appears to follow the spatial profile of the magnetic field, and to depend on its local strength for $V/t < (V/t)_c$. Near the zero-field criticality the profile of the OP follows the magnetic field's more closely, whereas at large interactions the effect of inhomogeneous magnetic field  becomes irrelevant, leading to uniform condensation. In Fig.\ [\ref{rotsymm}] (red curves), we exhibited the spatial distribution of the OP in presence of a non-uniform magnetic field, applied on a lattice that preserves the $C_6$ symmetry of a hexagon. Our computation yields an interaction induced OP  as a function of space qualitatively similar to the assumed profile of the magnetic field itself. \\

\begin{figure}[t]
{\centering\resizebox*{80mm}{!}{\includegraphics{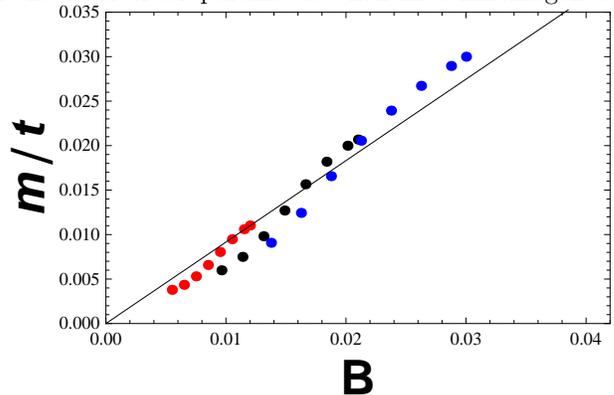}}
\par} \caption[] {The scaling of the local OP with local magnetic field (measured in the unit of $B_0$) at $V/t=0.05$ at various values of the total flux. Red, black and blue dots correspond to $\Phi_{total}=5.86 \Phi_0,10.14 \Phi_0$ and $14.6 \Phi_0$, respectively.
Different dots of the same color signify  local OP at various position in the bulk.}
\label{linear}
\end{figure}

Let us now turn to functional dependence of the OP on the magnetic field and interaction, when the former is
space dependent. At sufficiently weak interactions, the size of the gap at different region of the bulk of the system varies almost precisely linearly with the \emph{local} strength of magnetic field. The linear dependence of the local OP at $V/t=0.05 (\ll (V/t)_c)$ with the local magnetic field is shown in Fig. [\ref{linear}]. As the interaction is increased there is a
crossover to a sub-linear dependence of the mass on the local magnetic field. Situation is quite similar to the one
in presence of a uniform field. Right at the zero-field criticality $(V/t=0.75)$, the local OPs in the entire bulk of the system varies as $\sqrt{B}$, independent of the position (blue dots in Fig.\ [\ref{LLS}]). This suggests that the  OP in the insulating phase may be a universal function of the {\it local} magnetic field, independent of its spatial distribution. \\

\section{Summary and Discussion}

 In the present work we systematically studied magnetic catalysis for the spinless interacting electrons on a honeycomb lattice of finite extension, both for uniform and spatially modulated magnetic fields. In presence of the magnetic field, either uniform or non-uniform, the semimetal-insulator transition takes place at weak interaction in a large system. We here considered only the nearest-neighbor component ($V$) of the Coulomb interaction, and omitted its long-ranged $(\sim 1/r)$ tail \cite{gonzales, semenoff} for simplicity. We computed the self-consistent Hartree solution of the interaction-induced gap while keeping the system at the filling one half and
presented a scaling behavior of the interaction-induced order parameter (or a gap $(m)$) with the magnetic field and interaction, at $T=0$.
At weak interaction we observed a linear variation of the interaction induced local OP with the local magnetic field. With increase in the  strength of the interaction we find a crossover from linear to a sub-linear dependence of the mass on the local magnetic field. A perfect $\sqrt{B}$ dependence of the OP emerges when the system is tuned to be precisely at the zero-field   criticality, which we identified to be at $V/t=0.75$. This is close to the value found analytically \cite{Franz}.

In our analysis we have considered only the nearest-neighbor hopping amplitude $t$, while neglecting the next-nearest-neighbor hopping $t'$, which in graphene, for example, is finite but rather small. The main effect of a finite $t'$ is the violation of the perfect particle-hole symmetry of the free electron spectrum. On the basis of continuum theory we expect, however, that the inclusion of $t'$ would not change our results in a significant way, once the chemical potential is adjusted so that the central (formerly zero-energy) LL is half filled. A more detailed analysis is left for future study. 

If we were to restore the spin of electrons, we would need to include a finite on-site Hubbard interaction as well. In absence of a magnetic field, anti-ferromagnetic (AF) ground state, is energetically favored for a large on-site Hubbard interaction when the chemical potential is at the Dirac point \cite{bitan}. The presence of magnetic field stabilizes such ground state even at an infinitesimal on-site interaction $(U)$ \cite{herbut4, Meng}. We therefore expect that the system would develop a local expectation value of the N$\acute{e}$el order parameter when in a inhomogeneous magnetic field, if $U\gg V$. If $V\gg U$, on the other hand, the system would decrease  the energy more by forming a charge density wave order of the type we considered here. At zero magnetic field the two quantum phase transitions  belong to distinct Gross-Neveu universality classes \cite{vladimir, subir}. A scaling behavior in a similar model has also been studied recently in presence of both real and pseudo magnetic fields.\cite{bitanown}
\\

\section{ACKNOWLEDGEMENT}

This work is supported by the NSERC of Canada. B. Roy thanks Peter Smith and Kamran Kaveh for useful discussions, and to Vladimir Juri\v{c}i\'{c}, Kelly Cheng and Payam Mousavi for their help with the manuscript.

\end{document}